\documentclass[10pt,twocolumn,letterpaper]{article}
\pdfoutput=1 
\usepackage{icb}
\usepackage{times}
\usepackage{epsfig}
\usepackage{url}
\usepackage{graphicx}
\usepackage{amsmath}
\usepackage{textcomp}
\usepackage{amssymb}
\usepackage[export]{adjustbox}[2011/08/13]
\usepackage{multirow}
\graphicspath{{./Figures/}}
\usepackage{epstopdf}
\usepackage{caption}
\usepackage[bottom]{footmisc}
\usepackage{placeins}
\usepackage{enumitem}
\setlist[itemize]{leftmargin=*}
\usepackage{fancyhdr}

\usepackage[pagebackref=true,breaklinks=true,letterpaper=true,colorlinks,bookmarks=false]{hyperref}

\icbfinalcopy 
\ificbfinal\fi
\begin{document}
\hypersetup{pdfauthor={Umang Yadav, Sherif N Abbas, Dimitrios Hatzinakos},pdftitle={Evaluation of PPG Biometrics for Authentication in different states},pdfsubject={Evaluation of PPG Biometrics for Authentication in different states},pdfkeywords={PPG,Biometrics Authentication,Photoplethysmograph,Medical biometrics, PPG Dataset, BioSec.Lab}}
\title{Evaluation of PPG Biometrics for Authentication in different states}

\author{Umang Yadav \qquad
Sherif N. Abbas \qquad
Dimitrios Hatzinakos \\
The Edward S. Rogers Sr. Department of Electrical and Computer Engineering,
University of Toronto\\
10 Kings College Road, Toronto, ON, Canada, M5S 3G4.\\
{\tt\small \{umang,sseha,dimitris\}@ece.utoronto.ca}}


\maketitle
\thispagestyle{fancyplain}

\begin{abstract}
Amongst all medical biometric traits, Photoplethysmograph (PPG) is the easiest to acquire. PPG records the blood volume change with just combination of Light Emitting Diode and Photodiode from any part of the body. With IoT and smart homes' penetration, PPG recording can easily be integrated with other vital wearable devices. PPG represents peculiarity of hemodynamics and cardiovascular system for each individual. This paper presents non-fiducial method for PPG based biometric authentication. Being a physiological signal, PPG signal alters with physical/mental stress and time. For robustness, these variations cannot be ignored. While, most of the previous works focused only on single session, this paper demonstrates extensive performance evaluation of PPG biometrics against single session data, different emotions, physical exercise and time-lapse using Continuous Wavelet Transform (CWT) and Direct Linear Discriminant Analysis (DLDA). When evaluated on different states and datasets, equal error rate (EER) of $0.5\%$-$6\%$ was achieved for $45$-$60$s  average training time. Our CWT/DLDA based technique outperformed all other dimensionality reduction techniques and previous work.
\end{abstract}

\section{Introduction}

\begin{figure}[t]
\centering
\begin{minipage}{.20\textwidth}
  \centering
  \includegraphics[height=2cm,width=0.6\linewidth]{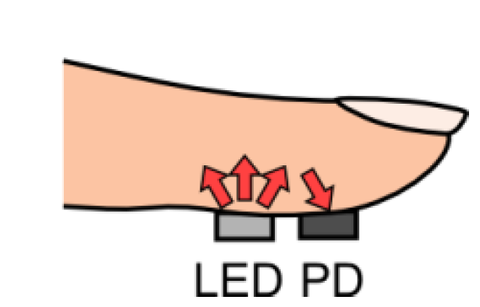}
  \captionof{figure}{PPG recording \newline with LED and Photo-Diode (PD)}
  \label{fig:ppg_led}
\end{minipage}%
\begin{minipage}{.30\textwidth}
  \centering
   \adjustbox{trim={.0\width} {.00\height} {0.00\width} {.00\height},clip}%
 {\includegraphics[width=0.9\textwidth, height=0.75\textwidth]{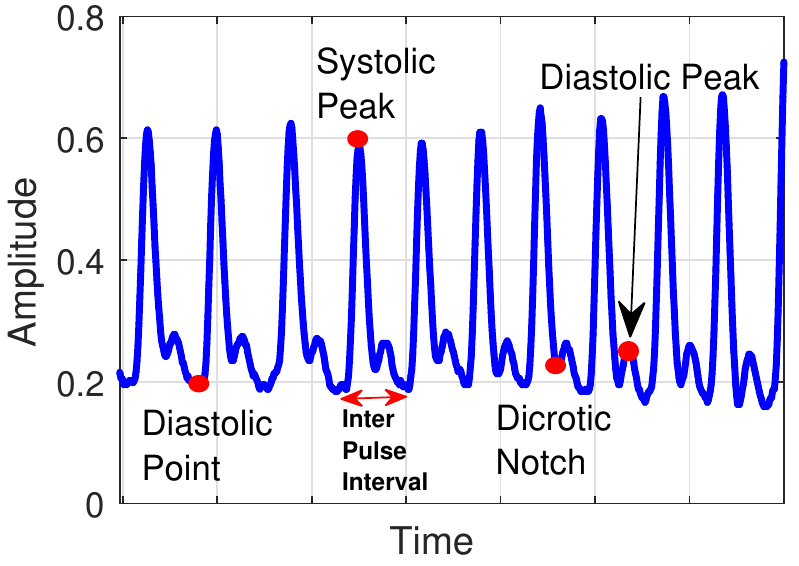}}%
 \captionof{figure}{Sample PPG Signal}
  \label{fig:ppg_sig}
\end{minipage}
\end{figure}

Risk of impersonation and violation in security of any system is ever lasting threat.~This risk can cause financial ruins and, is also life threatening in today's Internet of Things (IoT) and Smart Home Era. Hence, need of low cost and non-invasive authentication system has never been greater. This is why in last two decades, biometrics research has emerged as very important area of research. Photoplethysmograph (PPG) is a non-invasive electro-optical method which measures the volume of the blood flowing through the body part under testing. It reflects the pulsative actions of the arteries through the interaction of the oxygenized hemoglobin and photons. It is believed that every person has unique hemodynamics and cardiovascular system. Since, PPG captures this unique characteristics, this paper presents method to leverage it for biometric authentication. 
PPG being a biological signal, it is harder to steal or replicate. It has advantages of inherent anti-spoofing and liveness detection over traditional biometrics modalities. Also, as shown in Fig.\ref{fig:ppg_led}, PPG can be recorded with just combination of LED and Photo-Diode (PD) from any part of the body, which provides greater flexibility for systems design. Since, PPG recording only requires LED and PD, it is very cost effective compared to other biometric traits. In context of medical biometrics, PPG recording doesn't require any kind of gel (EEG), external stimulus (TEOAE) or multiple electrodes (ECG) and can be conveniently recorded from virtually any part of body. All such factors give PPG biometrics extra edge on other medical biometrics traits. 

Apart from authentication, PPG can be used for many clinical applications such as measuring oxygen saturation, blood pressure, detecting peripheral vascular diseases  etc. Wearable device companies such as Fitbit\textsuperscript{\tiny{TM}} are already using PPG signals for Heart Rate(HR) measurements. PPG based biometric system can be integrated into such devices and smart homes' IoT. Thus, PPG has inherent advantage of being portable and can give user unobtrusive experience. 

In practice, PPG signal is generally impaired by many common noise sources during acquisition such as motion artifacts, sensor movements, respiration, premature ventricular contraction (PVC) and ambient light. In addition, since PPG embodies proper functioning of heart, under physical and emotional stress, it changes along with Heart Rate (HR). Empirical comparison of PPG signal under rest condition and exercise shows considerable change in shape of PPG signal and its spectrum. Although indiscernible, it is also important to assess system performance under different emotions as they influence functioning of autonomic nervous system and heart. Furthermore, like any other biometric traits, it is crucial to take into account long-term behavior with time-lapse. In this paper, system performance is evaluated against all such variations using $3$ different datasets.  

For PPG based authentication, many fiducial and non-fiducial based approaches have been proposed in past. Fiducial based methods rely on detecting fiducial points of PPG signals shown in Fig.\ref{fig:ppg_sig} such as systolic peak, diastolic peak, dicrotic notch, inter-pulse interval, amplitudes of peaks, etc. Given variability in PPG shape in different states, fiducial detection on raw PPG signal might be unsuccessful or incorrect. Therefore, many researchers extended the idea to first derivative (FD) and second derivatives (SD) of raw PPG signals and used similar points on FD and SD as features for authentication.  For example, Kavsaoglu et.al. extracted 40 features from raw PPG, its FD and SD with kNN showed $94.44\%$ accuracy \cite{kavsaouglu}. On non-fiducial side, Spachos et.al. showed feasibility of PPG as biometrics identifier reporting $0.5\%$ and $25\%$ EER on two different datasets using temporal features and LDA \cite{spachos}. In most recent study, Karimian et.al. showed non-fiducial approach with DWT and kNN, reported $1.31\%$ mean EER for $42$ subjects \cite{karimian2017non}. Sarkar et.al. evaluated variations with emotions using DEAP dataset by approximating PPG beat with sum of gaussians and showed accuracy of $90.53\%$ for cross-emotions evaluation \cite{sarkar2016biometric}.

Many of the previous methods have been focused on fiducial approach. But fiducial points detection in every condition is error-prone. Therefore, in this paper non-fiducial approach is adopted. Also, most of the prior work has been focused on single session evaluation. For robustness, it is important to check variability in different states.  We present Continous Wavelet Transform (CWT) and Direct Linear Discriminant Analysis (DLDA) based method for authentication. CWT reveals unique time-frequency behaviour of each individual while DLDA selects CWT coefficients that give maximum class separabillity. Such feature selection is necessary for better results against variations. Additionally, taking more pragmatic approach and training of model was limited to $45-60$ seconds of signal and system performance was evaluated for different length of authentication time. 

Section \ref{sec2} and \ref{Other} present methodology and detailed description of system comprising pre-processing, template creation and template matching. Section \ref{results} presents experimental results in different emotional states, exercise, rest condition and under time lapse. Section \ref{conclusion} concludes the paper with observations and future work.

\section{PPG Authentication System and \newline Methodology} \label{sec2}
System flow is similar to many biometrics authentication system. As shown in Fig.\ref{flow} system consists of the three main blocks; pre-processing, template creation and template matching.
\subsection{Pre-processing}
\begin{itemize}[noitemsep]

\item \textbf{Filtering}: Butterworth IIR Band pass filter of cutoff frequencies $0.5$-$5$ Hz and order $38$ has been applied on whole signal for the removal of the power line interference, motion artifacts, baseline wanders.\ After filtering amplitude was normalized to have dynamic range of one. 

\begin{figure}[t]
\centering
\includegraphics[height=4cm,width=0.65\linewidth, trim=2.5cm 20cm 10cm 1cm]{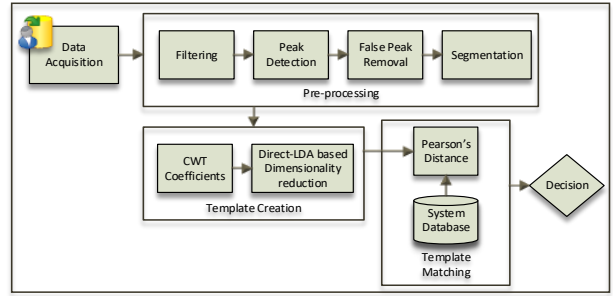}
\caption{System Flow Diagram}
\label{flow}
\end{figure}
 
\item \textbf{Peak Detection}: Successful application of pattern recognition techniques depends on reliable peak detection. Unlike ECG, there is no standard proven algorithm for systolic peak detection in PPG. In practice, most of the local maxima finding algorithms work reasonably well with PPG. To find peak locations, signal is first squared to emphasize amplitude difference between local maxima and other points. Thereafter, peaks were detected based on amplitude prominence. 

\item \textbf{False Peak Removal}: Many times motion artifacts, respiration, ambient light falls into passband of filter. In such cases, filtering doesn't help. Also, filtering degrades the amplitude of PPG peaks and causes false peak detections. To tackle this issue, simple false peak removal technique based on distance was employed. For any healthy subject based on his/her age and physical activity state if heart rate (HR) found using distance between two peaks was outside of normal range of HR, then those peaks were removed. 
 
\item \textbf{Segmentation}: For the features extraction, signal is segmented around detected peaks. Length of each segments is set using median HR found from signal. For given peak location $\textit{i}$, segment can be represented as follows:
\begin{equation}
S(j,1.....2r+1)= SIG(i-2r....i+2r)
\end{equation}
where r is median HR calculated from signal, \textit{j} is peak number, \textit{SIG} is signal and \textit{S} is segment. Since, each pulse segment covers, 3.5 to 4 pulses, it is larger in length and gives better frequency resolution with CWT. Hence, this choice of length was found to be producing good results. Furthermore, each two consecutive segments in $\textit{S}$ were averaged. This was necessary to improve SNR, damping any unwanted random noise as segment size was larger. 
\end{itemize}

\subsection{Template Creation}
\begin{itemize}[noitemsep]

\item \textbf{Features Extraction}:
CWT has been selected for the features extraction. Unlike discrete wavelet transform, CWT gives finer resolution without skipping samples. CWT reveals unique details of time-frequency variations in a single person. In time domain, Wavelet transform of signal is given by

\begin{equation}
WT_x(a,b)= \frac{1}{\sqrt{a}} \int\limits_{-\infty}^{\infty} x(t) .\ \psi^*\left(\frac{t-b}{a}\right)\mathrm{dt}
\end{equation}

In our case, $x(t)$ is a PPG segment. $\psi(t)$ is mother wavelet which is basically bandpass filter that changes with scale factor $a$ and translation $b$. By such translation and scaling, mother wavelet separates mixed frequency components of given PPG segment. Since analytic wavelets are more suitable for oscillatory signal and morse wavelet can approximate many other analytic wavelets by selecting appropriate parameters, in experiments \lq{Analytic Morse}\rq~wavelet was used as mother wavelet. 

\item \textbf{Dimensionality Reduction}: Linear discriminant analysis (LDA) based dimensionality reduction was applied on features vectors generated in previous stage. CWT along with LDA emphasizes time-frequency behavior in such a way that it increases the inter subject variability and reduces the intra subject variability. Such features selection is also important because PPG being physiological signal changes with time. Prior to dimensionality reduction, features were normalized on time axis using zero padding.
 
Given a training set $\mathbb{Z}$ = $ \left\{\mathbb{Z}_j \right\}_{j=1}^{K}$ containing \textit{K} classes each having $\mathbb{Z}_k$=$\{\textbf{z}_{ki}\}_{i=1}^{N_k}$ where $\textit{N}_k$ is number of feature vectors for each class \textit{k}, LDA finds projection weight $\mathbb{\textit{W}}$ that maximizes the fisher criterion function $\mathbb{J}(W)$,
\begin{equation}
\mathbb{J}(W)=arg \max_{W} \frac{| W^T \mathbb{S}_b W |}{|W^T \mathbb{S}_w W|}
\end{equation}
here $\mathbb{S}_b$ and $\mathbb{S}_w$ are the between class and within class scatter matrix respectively defined as,
\begin{equation}
\mathbb{S}_b= \sum_{k=1}^{K} N_k (\mu_k - \overline\mu) (\mu_k - \overline{\mu})^T 
\end{equation}
\begin{equation}
\mathbb{S}_w= \sum_{k=1}^{K} \sum_{i=1}^{N_k}(\textbf{z}_{ki} - \mu_k) (\textbf{z}_{ki} - \mu_k)^T
\end{equation}
where  $\mu_k $=$\frac{1}{N_k} \sum_{i=1}^{N_k} \textbf{z}_{ki}$ is mean of class $\mathbb{Z}_k$ and $\overline{\mu}$ is overall mean vector. LDA finds $\mathbb{\textit{W}}$ as $\textit{m}$=$K$-$1$ most significant eignevectors of $(\mathbb{S}_w)^{-1}(\mathbb{S}_b)$ that corresponds to first \textit{m} largest eigenvalues. After obtaining weight $\mathbb{\textit{W}}$, all the input training PPG segments are subjected to linear projection of $\tilde{\textbf{z}}=W^T\textbf{z}$. In practice, number of features might be larger than number of training samples. In that case, LDA suffers from small sample size problem. We used Direct-LDA (DLDA)  to address this issue\cite{yu2001direct}. Projected training vectors $\tilde{\textbf{z}}$ together with weight $\mathbb{\textit{W}}$ is saved in gallery for template matching.
\end{itemize}

\subsection{Template Matching}
Prior to template matching of test data, it is passed through all the blocks i.e. pre-processing and template creation. Weights of the LDA is pre-saved from the training session and test feature vectors are projected in similar way.  Template matching is carried out using Pearson's distance between training templates and test vectors. Pearson's distance $S_p$ between two vectors $a$ and $b$ is defined as:
\begin{equation}
S_p(a,b)=1-\frac{cov(a,b)}{\sqrt{cov(a,a) . cov(b,b)}}  
\end{equation}%
where $cov(a,b)$ is the covariance matrix of $a$ and $b$. Test vector is accepted if $S_p\leq threshold$. 

\begin{table*}[t] 
\caption{Capnobase dataset results with DLDA when evaluated with different number of randomly selected consecutive test segments (\textit{nTest}) and trained using only first 45 seconds.}
\label{capnobase_dlda45}
\setlength\tabcolsep{5.5pt}
\noindent\makebox[\textwidth]{%
\begin{tabular}{c|c|c|c|c|c|c|c|c|c}
\hline
\multicolumn{1}{c||}{\textit{nTest}} & 2 & 5 & 10 & 20 & 30 & 40 & 50  & 100 & All \\
\hline
\multicolumn{1}{c||}{mean EER} & 1.12\% & 1.09\% & 0.96\% & 0.85\% & 0.80\% & 0.80\%& 0.74\%& 0.58\% &0.46\%\\
\hline
\multicolumn{1}{c||}{std. EER} &0.44\%&0.45\%&0.38\%&0.29\%&0.27\%&0.25\%&0.21\%&0.14\% & 0.00\%\\
\hline
\end{tabular}}
\end{table*}

\begin{table*}[t] 
\centering
\caption{Capnobase dataset results with different methods when evaluated with All test segments. Here std.EER is zero.}
\label{capnobase_all}
\setlength\tabcolsep{4.5pt}
\begin{tabular}{c|c|c|c|c|c|c|c|c}
\hline
\multicolumn{1}{c||}{Method} & CWT/DLDA & CWT/KDDA & Openset & CWT/KPCA & CWT/LDA & CWT/PCA & Karimian et.al.  & AC/LDA \\
\hline
\multicolumn{1}{c||}{EER} & 0.46\% &2.32\% & 2.50\% &2.38\% & 2.32\%& 4.01\%&1.51$\pm$0.48\% &4.82\%\\
\hline
\end{tabular}
\end{table*}

\section{Other Implemented Methods}\label{Other}
Apart from DLDA, few other techniques were also implemented to reproduce and compare the results. 
\begin{itemize}[noitemsep]

\item \textbf{DWT/kNN}: It is based on method presented in \cite{karimian2017non} by Karimian et.al. It is a non-fiducial method which uses DWT coefficients extracted using coiflet wavelet from PPG segment as features. It implements two steps process based on kolmogrov smirnov based correlation filter (ksCBF) and Kernel PCA (KPCA) to remove correlated features and to reduce dimensionality. Classification is carried out using local density factor based kNN. 

\item \textbf{AC/LDA} \cite{aclda}:
It is widely used in ECG based recognition system. It doesn't require peak detection in PPG. Instead, signal is blindly segmented into overlapping windows of predefined length. Then, normalized auto-correlation (AC) of each window is calculated as following:
\begin{equation}
\hat{R_{xx}}[m]=\frac{\sum_{i=0}^{N-|m|-1}x[i]~x[i+m]}{\hat{R_{xx}}[0]}
\end{equation}
where, $x[i]$ represents windowed PPG singal, $x[i+m]$ delayed PPG singal with time lag of $m=0,1,2.....M-1$ and $M<<N$. 	These AC windows are projected on lower dimensional space using LDA. Finally template matching is done with euclidean distance. 

\item \textbf{Openset Validation}:
It follows same system as decribed in Fig.\ref{flow} except doing dimensionality reduction. In openset validation, CWT PPG feature segments are stored in gallery without subjecting them to dimensionality reduction. During template matching, same process is followed and CWT features of claimed person is compared against CWT feature segments stored in gallery without projecting them to any lower dimensional space. 

\item \textbf{Other Subspace Learning Techniques}: Apart from DLDA other subspace learning techniques such as PCA, LDA, KPCA and Kernel Direct Discriminant Analysis (KDDA) \cite{KDDA} were also implemented. For all of these techniques same system as Fig.\ref{flow} was followed with DLDA dimensionality block replaced with different technique. For both KPCA and KDDA, Gaussian kernel was used. Reduced dimensionality of all these techniques was set to $N-1$, where $N$ is number of subjects in dataset.
 
\end{itemize}

\section{Experimental Results and Discussion}\label{results}

\subsection{Performance Metric}

The performance of all techniques was evaluated in verification mode (1 to 1 matching). In this configuration, subject can claim an enrolled ID and based on matching score system rejects or accepts the claim. In this setting, error can be characterized by False Rejection Rate (FRR) and False Acceptance Rate (FAR). ROC (Receiver Operating Characteristic) curve plots FRR vs FAR for different operating points. Operating points are matching score or threshold values. Trade off appears between FAR and FRR. EER (Equal Error Rate) is an operating point on ROC where FAR=FRR. For experiments, EER is chosen as a metric to compare different techniques. In practice, choice of operating point is application dependent.

\subsection{Results}
In experiments, evaluation strategy was designed such that it addressed the issues of robustness against different stress conditions, against time and speed. To assess the system performance against all such conditions, three different datasets were used. All of three datasets contained data recorded in single session, in different emotional stress, physical stress and data recorded after time-lapse. In experiments, we ran multiple simulations varying different parameters such as CWT scale, segment length, distance metric etc.  It was also found that using larger segment length gave better results in general, since CWT gave better frequency resolution. Segment length was set based on median HR such that it covered 3.5-4 cycles of PPG. In addition, two consecutive time segments were averaged. This reduced variance due to noise in larger segment and across whole period for one subject.  Experimental results also supported this hypothesis.  One more engineering problem was to select number of scales from CWT to maximize discrimination while keeping feature vector small. It was found that choosing coefficients from only one scale, which filters signal between 1-2 Hz gave better results. 1-2 Hz is also most prominent frequency range of PPG signal. However due to diverse nature and different sampling frequency of each of three dataset, best CWT scale for classification was different for each of the three dataset. 

Another challenge is of speed. PPG is slower signal compared to traditional biometric traits and requires larger training and testing time. In experiments taking more pragmatic approach, average training time was limited between $45-60$s. On the other side to assess effect of limited authentication time, different number of test segments were selected to vary available test time. Experimental results for each of three dataset under various configurations is presented in subsequent discussions. 

\begin{figure}[t]
\centering
\adjustbox{trim={.00\width} {.00\height} {0.00\width} {0.00\height},clip}%
 {\includegraphics[width=0.45\textwidth, height=0.32\textheight]{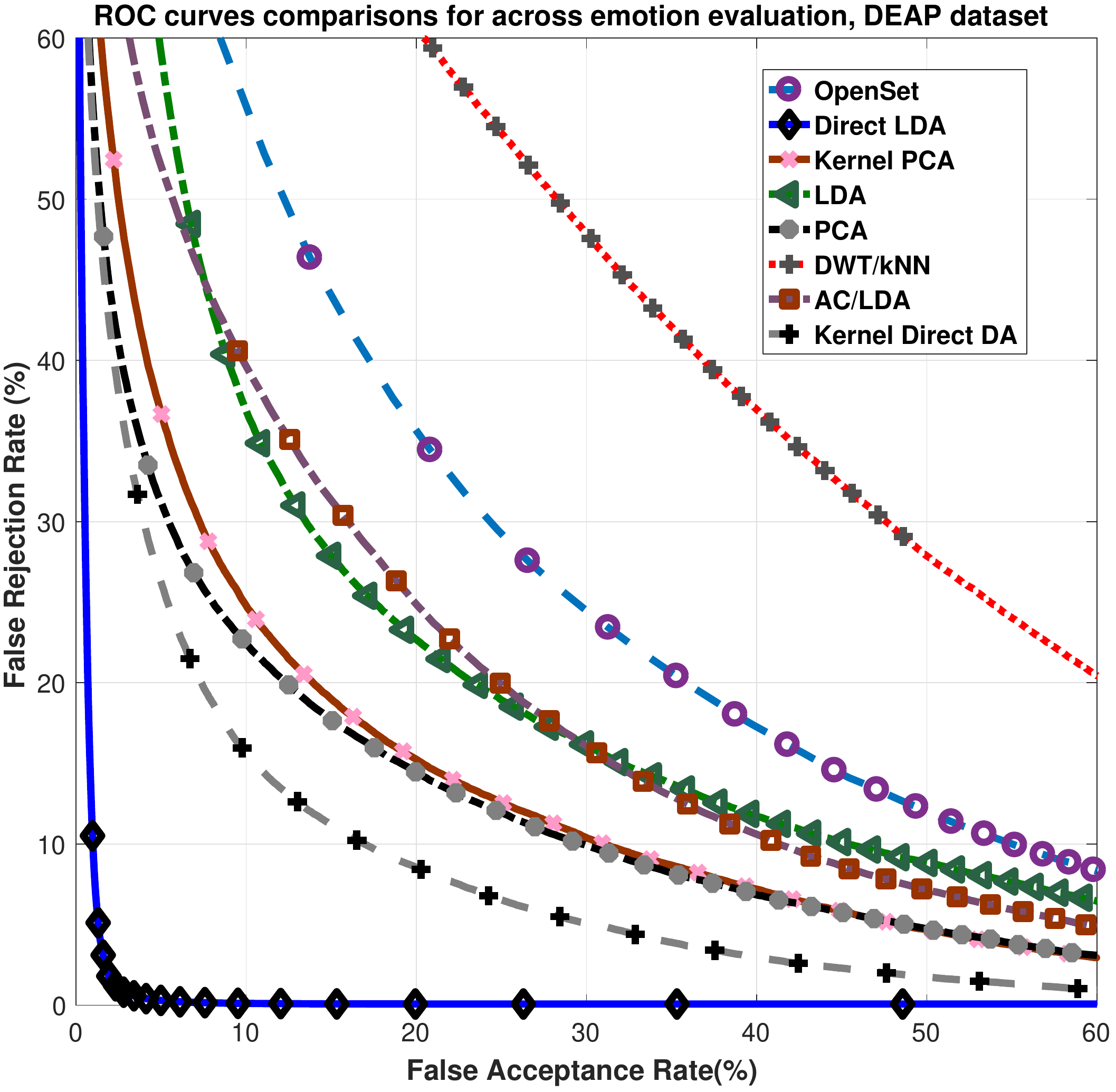}}
\caption{DEAP dataset ROC Curves}
\label{DEAP:ROC_curves}%
\end{figure} 
Capnobase dataset was used for single session evaluation \cite{capnobase}. It consists of $8$ minute single session data recorded in relax condition from $42$ subjects. Training time was fixed to first $45$s of signal, while different number of consecutive test segments (\textit{nTest}) were selected randomly over 50 iterations. EER was calculated for each iteration. To select best scale, numerous experiments were carried out varying training time, \textit{nTest} and CWT Scale. Based on these experiments $3rd$ highest scale was chosen for further experiments. Similar experiments were also carried out to select best scale for other two datasets.  Results in Tab.\ref{capnobase_dlda45} and \ref{capnobase_all} present mean EER and standard deviation in EER (std. EER) over 50 iterations for training time of 45 seconds only. CWT/DLDA performed better than every other technique. During the experiments, it was found the increasing training time decreased EER. However, drop wasn't significant. Also, increasing testing time (\textit{nTest}) also decreased EER. Nevertheless, EER of $1.12\pm0.44\%$ was achieved using $45$s of training and just $2$ segments which required $6$-$7$s in time. 

Next, system was tested for emotional robustness using DEAP Dataset \cite{DEAP}. DEAP dataset consists of data recorded in $40$ different kinds of emotions spread over valence arousal plane for $32$ subjects. Data was recorded on same day with short breaks and baseline emotional stimuli in between. For each subject, model was trained using data recorded in one type of emotion, that is total one minute of training of data and tested it against rest of the dataset. For each subject, dataset consisted of $39$ genuine trials and $1240$ (40 data samples from rest of the $31$ subjects) imposter trials. EER was calculated over 40 iteration, each time model trained with different emotion. Here, $nTest$ were not selected randomly, as test data were from different sessions. Tab.\ref{DEAP_dlda} presents mean EER and std. EER for DLDA. Remarkably, EER of 2.61$\pm$1.14 \% was achieved using 60s of training time and 2 test segments or $6$-$7$s of test time. Fig.\ref{DEAP:ROC_curves} presents comparison of different techniques using ROC curves. It can be noticed that all other techniques performed poorly compared to DLDA. Also CWT/DLDA produced better result than \cite{sarkar2016biometric} on same dataset.

\begin{figure*}[t]
\centering
\begin{minipage}[b]{.50\textwidth}%
  \centering
  \adjustbox{trim={.0\width} {.\height} {0.\width} {0.0\height},clip}%
 {\includegraphics[width=1.\textwidth, height=0.23\textheight]{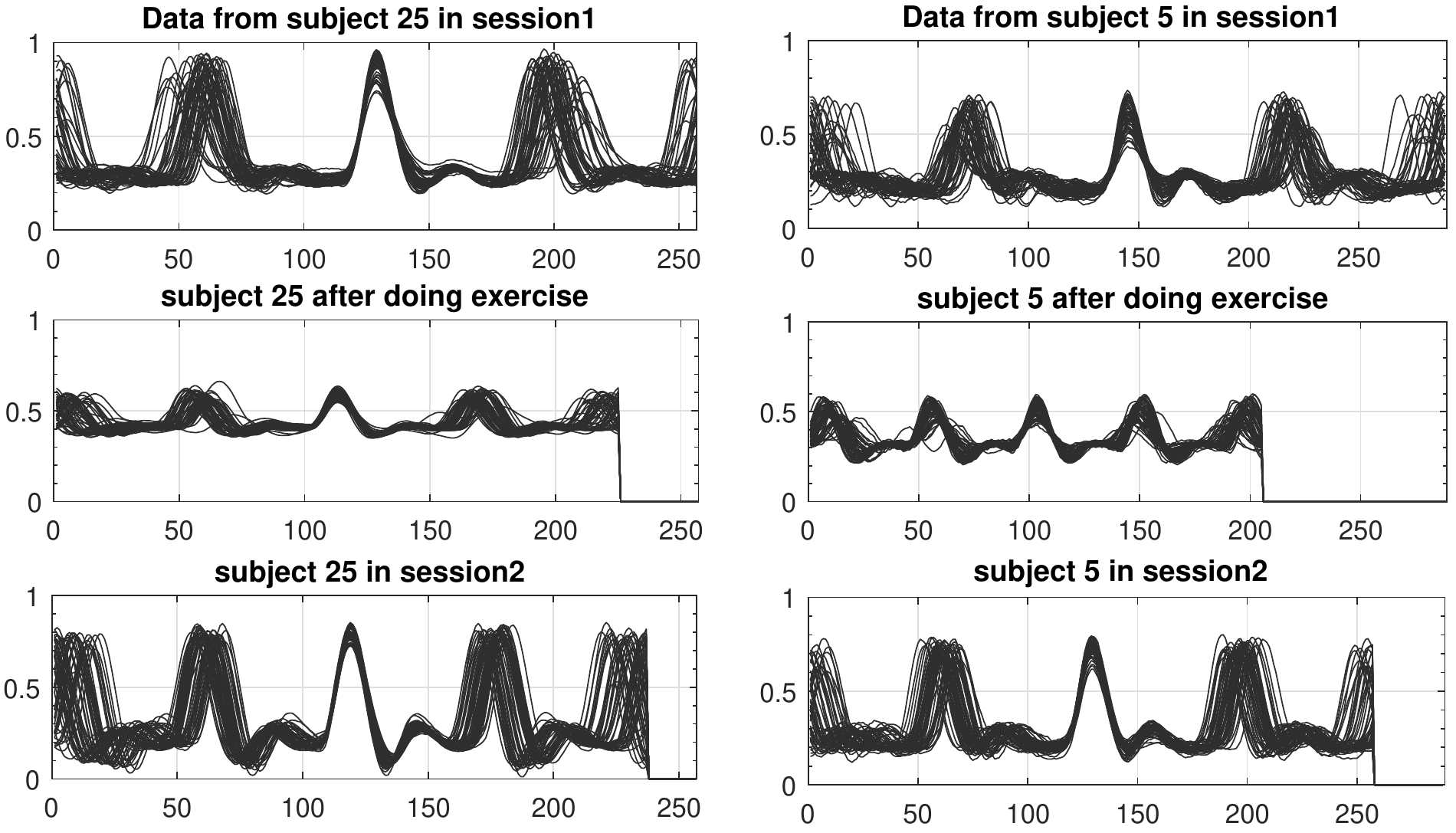}}
 \captionof{figure}{Variations in PPG signal with exercise and time-lapse for BioSec.Lab PPG dataset}
 \label{Bioseclab:variations}
\end{minipage}%
\begin{minipage}[b]{0.50\textwidth}%
\centering
\adjustbox{trim={.00\width} {.00\height} {0.0\width} {0.00\height},clip}%
 {\includegraphics[width=1.05\textwidth, height=0.25\textheight]{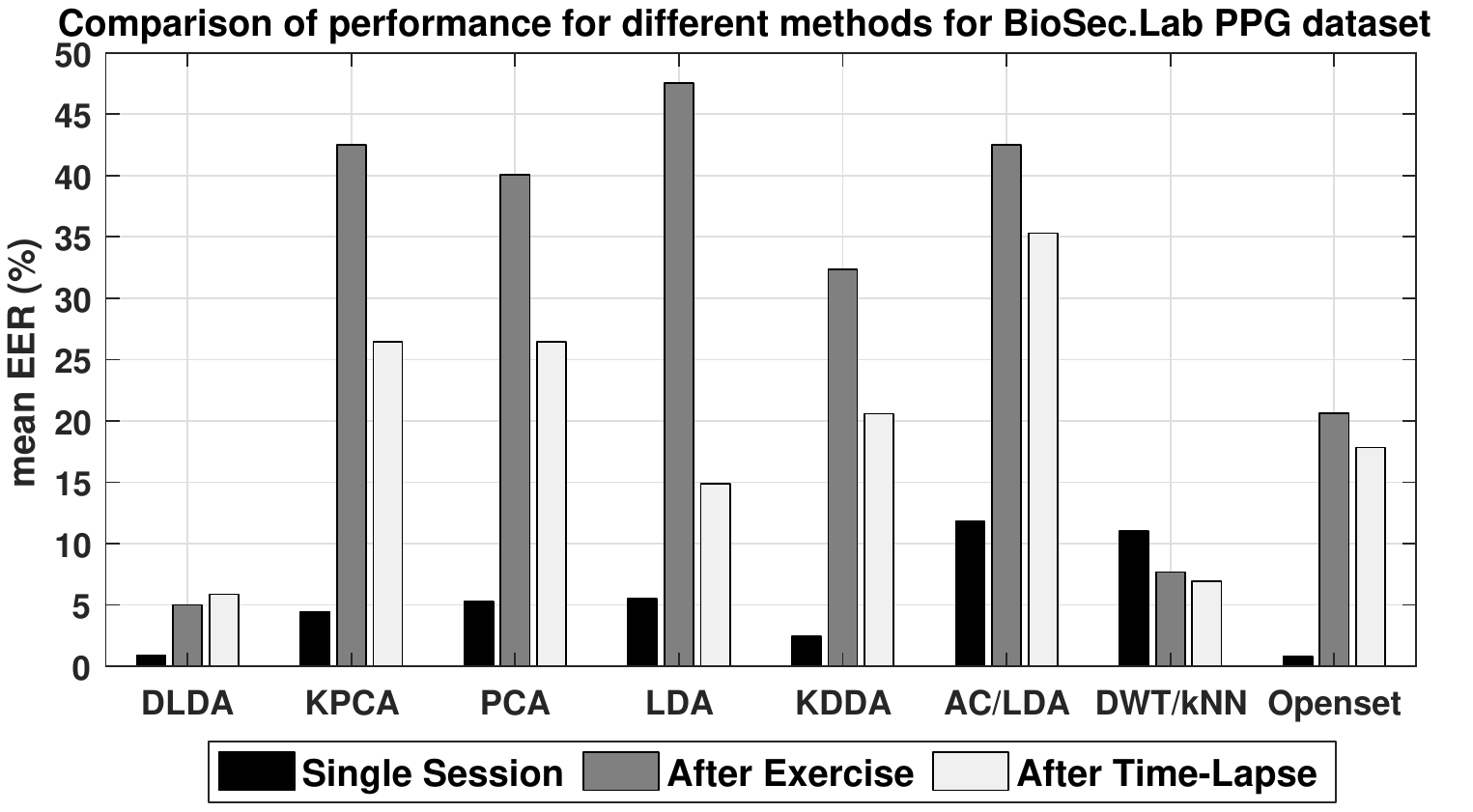}}
\captionof{figure}{Comparison of EER in different cases}
\label{Bar_plot} 
\end{minipage}
\end{figure*}

\begin{table}[t] 
\centering
\caption{Across session evaluation of DEAP Dataset using DLDA, \textit{nTest} is number of selected test segments}
\label{DEAP_dlda}
\setlength\tabcolsep{5.4pt}
\begin{tabular}{c||c|c|c|c|c}
\hline
\textit{nTest} & 2 & 5 & 10 & 20 & All \\
\hline
Mean EER & 2.61\% & 2.30\% & 2.13\% & 2.00\% & 2.11\% \\
\hline
std. EER & 1.14\% & 0.97\% & 0.87\% & 0.82\% & 0.87\% \\ 
\hline
\end{tabular}
\end{table}
Apart from emotions, PPG signal is susceptible to variations due to exercise because of heart rate change. For robust system, these variations can not be overlooked. In addition, as shown in Fig.\ref{Bioseclab:variations}, PPG being physiological signal, changes with time. Therefore, consequences of ignoring it can be disastrous. Like all other biometrics traits, permanence test on PPG is important. Since, no large dataset is available in public to estimate the effect of physical stress and time-lapse, BioSec.Lab PPG dataset was developed at University of Toronto \cite{BSL}.  PPG signals were recorded from fingertip using Plux Sensor over two sessions \cite{plux}. First session was conducted in two parts. First, PPG was recorded in relax condition for 3 minutes. Then subjects were asked to perform some form of intense exercise such as climbing up the stairs very fast to increase HR. PPG signal was again recorded for 3 minutes just after exercise. In total $41$ subjects participated in first session. In second session, signals were only recorded in relax condition for 3 minutes.  Two sessions were separated atleast $2$ weeks apart in time. However only $34$ subjects out of $41$ participated in second session. For evaluation, Model was trained using only initial $45$ seconds of relax data from first session. To assess, effect of physical stress, model was then tested on exercise data with different $\textit{nTest}$ selected from the initial signal non-randomly. To estimate time-robustness, again model was trained using $45$ seconds of relax data from first session and then tested on session $2$ data non-randomly. Tab.\ref{BioSec_ex_dlda} presents results for each case where single session performance on $45$ seconds of training data is included for comparison. 

From Tab.\ref{BioSec_ex_dlda}, it can be observed that EER increased by $5\%$ because of time-lapse and exercise when tested with whole signal. However, from Fig.\ref{Bar_plot} it can be seen that, CWT/DLDA performed far better than any other technique. Surprisingly, DWT/kNN method, which performed very badly for DEAP dataset, had decreasing EER for BioSec.Lab dataset, across different sessions. This can be due to that fact that correlation based filter empolyed in this method does not take into account robustness of features across sessions while removing them. Hence it is possible that because of reduced number of features, DWT/kNN gave decreasing error across 2 sessions for BioSec.Lab dataset, but on average when tested across $40$ sessions in DEAP dataset, it performed worst.  Also, ideally EER should decrease with increase in \textit{nTest}. This trend is visible in Tab.\ref{capnobase_dlda45}. But in Tab.\ref{DEAP_dlda} and Tab.\ref{BioSec_ex_dlda}, it is not consistent. This is due to the quality of samples. For example in Tab.\ref{BioSec_ex_dlda} across exercise, \textit{nTest}$=20$ has lower EER compared to All test segments. Because, it is possible that whole test signal (all segments) would have more deviated samples under physical or mental stress compared to only subset (e.g. $20$ segments) of test 
signal.  
\begin{table}[t]
\centering
\caption{BioSec.Lab PPG Dataset results using CWT/DLDA using only 45s of training data, here \textit{nTest} is number of selected test segments.}	
\label{BioSec_ex_dlda} 
\setlength\tabcolsep{5.5pt}
\begin{tabular}{c||c|c|c|c|c}
\hline
\textit{nTest} & 2 & 5 & 10 & 20 & All \\
\hline
\hline
\multicolumn{6}{c}{\text{Single Session Evaluation}} \\
\hline
Mean EER & 1.05\% & 1.00\% & 0.97\% & 0.86\% & 0.86\% \\
\hline
std. EER & 0.22\% & 0.18\% & 0.17\% &0.02\% & 0.02\%\\ 
\hline 
\hline
\multicolumn{6}{c}{\text{Across Exercise Evaluation}} \\
\hline
\multicolumn{1}{c||}{EER} & 2.50\% & 2.50\% & 2.95\% & 2.50\% & 5.00\% \\
\hline 
\hline
\multicolumn{6}{c}{\text{Across Session Evaluation}} \\
\hline
\multicolumn{1}{c||	}{EER} & 6.86\% & 8.65\% & 8.82\% & 8.82\% & 5.88\% \\
\hline
\end{tabular}
\end{table}

\section{Conclusion}\label{conclusion}
In this paper, CWT/DLDA based method was presented for PPG based authentication. By nature, CWT reveals idiosyncratic time-frequency behavior of PPG for each individual.  To get holistic understanding, method was evaluated for different datasets under different conditions. It was found that, system performed better with larger training and testing time. However by considering more practical scenario training time was fixed to only 45-60s. Better results of our method compared to previous works is attributed to the fact that, analytic wavelet are more suitable for oscillatory signals compared to non-analytic and DLDA boosts system performance by only choosing features that maximizes discriminality in class specific way. It was found that, physical stress and time-lapse has adverse effect on system. In future, we plan to  address these issues with more robust feature selection, continuous authentication and training model under many different conditions. 
{\small
\bibliographystyle{ieee}
\bibliography{lniguide}
}

\end{document}